\documentclass{elsart}
\usepackage{amssymb}

\renewcommand{\bar}[1]{\overline{#1}}

\usepackage{indentfirst}
\usepackage{psfig,color}
\usepackage{epsfig}
\usepackage{epsf}
\usepackage{graphicx}

\journal{Physics Letters B}

\begin{document}

\begin{frontmatter}
\title{Neutrino oscillations in de Sitter space-time}

\author[pku]{Xin-Bing Huang \corauthref{cor}}
\corauth[cor]{Corresponding author.} \ead{huangxb@pku.edu.cn}
\address[pku]{Department of Physics, Peking University, Beijing 100871, China}

\begin{abstract}
We try to understand flavor oscillations and to develop the
formulae for describing neutrino oscillations in de Sitter
space-time. First, the covariant Dirac equation is investigated
under the conformally flat coordinates of de Sitter geometry.
Then, we obtain the exact solutions of the Dirac equation and
indicate the explicit form of the phase of wave function. Next,
the concise formulae for calculating the neutrino oscillation
probabilities in de Sitter space-time are given. Finally, The
difference between our formulae and the standard result in
Minkowski space-time is pointed out.
\end{abstract}

\end{frontmatter}

\par

\section{Introduction}

Neutrino oscillations have been the most natural explanation to
the so-called solar neutrino problem and atmospheric neutrino
problem~\cite{mv04}. Since neutrino oscillations may provide a
tool both to understand elementary particles and their
interactions and to uncover the secrets on ancient celestial
bodies, they have attracted a lot of attention in modern
physics~\cite{xin04}. In Minkowski space-time, neutrino
oscillations have been extensively studied in the past by using
both plane waves~\cite{pon57} and wave packets~\cite{kay81,gkl91}
to represent the emitted neutrinos. But, the neutrinos emitted by
ancient sources must travel a very long distance to arrive at our
detectors, and be affected by gravitation. Therefore, in more
recent years, physicists have turned their attention to
specifically gravitational contributions to neutrino oscillations.

According to the methods appeared in the literature, we divide
those considerations into three categories:

\begin{enumerate}

\item Using approximate operator~\cite{prw96,gl97,cf97,wud01}. In
these papers, the effective Hamiltonian or the semiclassical
approximation, such as, $p_{x}\simeq E-\frac{m^2}{2E}$, is
generally used to investigate the complicated physical system.
This kind of method is fit for considering matter effects on
neutrino oscillations.

\item Adopting semiclassical phase
factor~\cite{ab96,fgks97,ab98,bhm99}. This category of study first
appeared in Ref.\cite{ab96}. All of these papers laid their
foundation on an extended covariant phase factor, for instance,
$\Phi=\int_{A}^{B}p_{\mu}dx^{\mu}$, given in Ref.\cite{sto79}.
Since the semiclassical action is taken as the quantum phase,
couplings between spin and gravitation are thus ignored.

\item Studying the covariant Dirac equation~\cite{kk98}. It seems
that this is the unique way to acquire the exact formulae for
calculating the oscillation probabilities. But, only in several
ideal models, one can obtain the exact solutions of the Dirac
equation. So this method can be used in a few cases.
\end{enumerate}

Recent astronomical observations on Type Ia
supernova~\cite{sch98,rie98,per99} and the cosmological microwave
background radiation~\cite{ber00,jaf01,ben03} imply that dark
energy can not be zero, and which will dominate the doom of our
Universe. Therefore, our observable universe will be an
asymptotical de Sitter space-time. On the other hand, the
scenarios of inflation usually need an inflationary epoch, in
which space-time can be treated as a de Sitter
manifold~\cite{lr99}. Upon above considerations, we investigate
neutrino oscillations in de Sitter space-time in this paper.

The different evolution of the phases of mass eigenstates plays
the pivotal role in neutrino oscillations. In Minkowski
space-time, the phase of a free boson or of a free fermion is
given by solving the Klein-Gordon equation or Dirac equation, that
is, $\Phi(t,\vec{x})=-(E\cdot t-\vec{p}\cdot\vec{x})$. From this
point of view, we need to solve the covariant Dirac equation to
get the exact form of the phase of wave function. After that, we
can obtain the exact formulae for calculating the neutrino
oscillation probabilities.

This paper is organized as follows. The covariant Dirac equation
in de Sitter manifold is explicitly given in section 2. In section
3, the exact solutions of the Dirac equation are obtained. After
that, we read out the phase factor. We then present the formulae
on neutrino-oscillation probabilities in section 4. Conclusions
and remarks appear in the last section.

\section{Dirac equation in de Sitter space-time}

To set the stage for studying neutrino oscillations in a curved
space-time, we introduce the local tetrad field as
follows\footnote{We use Roman suffixes to refer to the bases of
local Minkowski frame, and use Greek suffixes to refer to
curvilinear coordinates of space-time.}
\begin{equation}
\label{tetrad1} {\rm
g}^{\mu\nu}=e^{\mu}_{~a}(x)e^{\nu}_{~b}(x)\eta^{ab}~,
\end{equation}
where ${\rm g}^{\mu\nu}$ being the space-time metric, and the
metric tensor of Minkowski space-time $\eta_{ab}$ satisfy
\begin{equation}
\label{metric102}
\eta^{00}=+1~,~~~~\eta^{11}=\eta^{22}=\eta^{33}=-1~,~~~~
\eta^{ab}=0~~~~{\rm for}~~~~a \neq b~.
\end{equation}

In a curved space-time, the covariant Dirac equation
reads\footnote{we adopt the conventions that $x^{0}=t$, $x^{1}=r$,
$x^{2}=\theta$, $x^{3}=\varphi$ and $\hbar=c=1$.}
\begin{equation}
\label{Diraceq} i\gamma^{\mu}(x)\left[ \frac{\partial}{\partial
x^{\mu}}- \Gamma_{\mu}(x)\right]\psi(x)=m\psi(x)~.
\end{equation}
Here $\gamma^{\mu}(x)$ are the curvature-dependent Dirac matrices
and $\Gamma_{\mu}(x)$ are the spin connections to be determined.
The curvature-dependent Dirac matrices $\gamma^{\mu}(x)$ are
presented in terms of the tetrad field as
\begin{equation}
\label{Diracmatrix} \gamma^{\mu}(x)=e^{\mu}_{~a}(x)\gamma^{a}~,
\end{equation}
where $\gamma^{a}$ denotes the standard flat-space Dirac matrices,
which satisfies $\{\gamma^{a},\gamma^{b}\}=2\eta^{ab}$.

Under a general coordinate transformation, $x^{\mu}\longrightarrow
y^{\nu}$, it is well known that the spinor keeps
invariant~\cite{bw57}, that is,
\begin{equation}
\label{31-2} \psi(y)=\psi(x)~.
\end{equation}
Hence the phase difference of two points for a free neutral
spin-$\frac{1}{2}$ particle is invariant under the coordinate
transformation. As we will demonstrate, the phenomenon of flavor
oscillation is only related with the phase difference and the
mixing angles, one then can select any convenient coordinates to
study flavor oscillation. Because of this property, we choose the
comoving coordinates to investigate the neutrino oscillations in
de Sitter space-time.

In the conformally flat coordinates the metric of de Sitter
space-time is taken to be
\begin{equation}
\label{RWmetric} ds^{2}={\rm
g}_{\mu\nu}dx^{\mu}dx^{\nu}=dt^{2}-R^{2}(t)\left[
dr^2+r^2(d\theta^2+\sin^2\theta d\varphi^2)\right]~,
\end{equation}
where $R(t)=e^{Ht}$ is called cosmic scale factor of de Sitter
space-time. $H$ is a constant.

The tetrad field in the case of de Sitter metric (\ref{RWmetric})
is thus taken to be
\begin{equation}
\label{tetrad2} e^{\mu}_{~a}(x)={\rm diag} \left(
1,\frac{1}{R},\frac{1}{R r},\frac{1}{R r\sin\theta} \right)~.
\end{equation}
Inserting Eq.(\ref{tetrad2}) into Eq.(\ref{Diracmatrix}) yields
\begin{eqnarray}\nonumber
\gamma^{0}(x)&=&\gamma_{0}~,~~~~~~~~~~~~~~\gamma^{1}(x)=-\frac{1}{R}\gamma_{1}~,
\\
\label{con} \gamma^{2}(x)&=&-\frac{1}{R
r}\gamma_{2}~,~~~~\gamma^{3}(x)=-\frac{1}{R
r\sin\theta}\gamma_{3}~.
\end{eqnarray}

The spin connections $\Gamma_{\mu}(x)$ in Eq.(\ref{Diraceq})
satisfy the equation
\begin{equation}
\label{spincon}
[\Gamma_{\mu}(x),\gamma^{\nu}(x)]=\frac{\partial\gamma^{\nu}(x)}{\partial
x^{\mu}}+\Gamma^{\nu}_{\mu\rho}\gamma^{\rho}(x)~,
\end{equation}
where $\Gamma^{\nu}_{\mu\rho}$ are the Christoffel symbols for the
metric (\ref{RWmetric}), which are determined by
\begin{equation}
\label{Christoffel} \Gamma^{\nu}_{\mu\rho}=\frac{1}{2}{\rm
g}^{\nu\tau} \left(\frac{\partial {\rm g}_{\tau\rho}}{\partial
x^{\mu}}+\frac{\partial {\rm g}_{\mu\tau}}{\partial
x^{\rho}}-\frac{\partial {\rm g}_{\mu\rho}}{\partial x^{\tau}}
\right)~.
\end{equation}
Hence Eq.(\ref{spincon}) can be solved for the spin connections
$\Gamma_{\mu}(x)$ which we determine as
\begin{eqnarray}\nonumber
\Gamma_{0}&=&0~,~~~~~~~\Gamma_{1}=\frac{\dot{R}}{2}\gamma_{0}\gamma_{1}~,
\\
\label{spinconnection}
\Gamma_{2}&=&\frac{1}{2}\dot{R}r\gamma_{0}\gamma_{2}+\frac{1}{2}\gamma_{2}\gamma_{1}~,
\\
\nonumber
\Gamma_{3}&=&\frac{1}{2}\dot{R}r\sin\theta\gamma_{0}\gamma_{3}
+\frac{1}{2}\sin\theta\gamma_{3}\gamma_{1}+\frac{1}{2}\cos\theta\gamma_{3}\gamma_{2}~,
\end{eqnarray}
where $\dot{R}$ denotes $\frac{d R(t)}{dt}$, so that the
combination $\gamma^{\mu}(x)\Gamma_{\mu}(x)$ in Eq.(\ref{Diraceq})
simplifies to
\begin{equation}
\label{spinconsum}
\gamma^{\mu}(x)\Gamma_{\mu}(x)=-\frac{3\dot{R}}{2R}\gamma_{0}+\frac{1}{R
r}\gamma_{1} +\frac{\cos\theta}{2R r\sin\theta}\gamma_{2}~.
\end{equation}

Inserting Eq.(\ref{con}) and Eq.(\ref{spinconsum}) into
Eq.(\ref{Diraceq}), then we obtain the Dirac equation in de Sitter
space-time
\begin{equation}\label{diraceq2}
\begin{array}{l}
\displaystyle\left\{ iR\gamma_{0} \left( \frac{\partial}{\partial
t}+\frac{3\dot{R}}{2R} \right) - i \left[ \gamma_{1}
\left(\frac{\partial}{\partial r}+\frac{1}{r} \right)
+\frac{1}{r}\gamma_{2}\left(\frac{\partial}{\partial\theta}+
\frac{\cos\theta}{2\sin\theta}
\right)\right.\right.\\[1cm]
\displaystyle~~~~~~~~~~~\left.\left.+\frac{1}{r\sin\theta}\gamma_{3}\frac{\partial}{\partial\varphi}
\right]- m R \right\} \psi(t,r,\theta,\varphi)=0~.
\end{array}
\end{equation}


\section{The exact solutions of the Dirac equation}
We can simplify the Dirac equation (\ref{diraceq2}) by
transforming~\cite{lg82,hua05}
\begin{equation}
\label{seperation}
\psi(t,r,\theta,\varphi)=(\sin\theta)^{-\frac{1}{2}}
R^{-\frac{3}{2}}\Psi(t,r,\theta,\varphi)~.
\end{equation}
The reduced equation in terms of $\Psi(t,r,\theta,\varphi)$ is of
the form
\begin{equation}
\label{neweq} R\left( \frac{\partial}{\partial t }+ i m \gamma_{0}
\right)\Psi=\left[\gamma_{0}\gamma_{1}\left(\frac{\partial}{\partial
r}+\frac{1}{ r}\right)+\frac{1}{
r}\gamma_{1}\hat{K}(\theta,\varphi)\right]\Psi~,
\end{equation}
where
\begin{equation}
\label{Kdef} \hat{K}(\theta,\varphi)=\gamma_{0}\gamma_{1}\left(
\gamma_{2}\frac{\partial}{\partial\theta}+\gamma_{3}\frac{1}{\sin\theta}\frac{\partial}{\partial\varphi}
\right)
\end{equation}
is a Hermitian operator, as we know, which is defined in
Ref.\cite{sch38} first. The eigen equation of $\hat{K}$ is
\begin{equation}
\label{Keigen} \hat{K}\Psi_{\varsigma,\kappa}=
\varsigma\Psi_{\varsigma,\kappa}~,~~~~~~~~\varsigma=0, \pm 1, \pm
2, \cdot\cdot\cdot,
\end{equation}
where the eigen value $\varsigma$ is an integer, which has been
proven in Ref.\cite{bw57,sch38}.

Only the two matrices $\gamma_{0}$ and $\gamma_{1}$ remain
explicitly in the simplified Dirac equation (\ref{neweq}), they
can therefore be represented by $2\times2$ matrices
\begin{equation}
\label{twocomponent} \gamma_{0}=\left(
\begin{array}{cc}
{\bf 1} & 0
\\
0& -{\bf 1}
\end{array}\right)~,~~~\gamma_{1}=\left(
\begin{array}{cc}
0 & -\tau_{1}
\\
\tau_{1} & 0
\end{array}\right)~,
\end{equation}
where ${\bf 1}$, $\tau_{1}$ are respectively $2\times 2$ unit
matrix and the Pauli matrix, namely
\begin{equation}
\label{paulima} \tau_{1}=\left(
\begin{array}{cc}
0 & 1
\\
1& 0
\end{array}\right)~.
\end{equation}
We separate the angular factor from the wave function, and
represent the radial and temporal factor by a two-component
spinor, that is
\begin{equation}\label{sep}
\Psi_{\varsigma,\kappa}=\Theta_{\varsigma}(\theta,\varphi) \left(
\begin{array}{c}
\phi(r,t)
\\
\chi(r,t)
\end{array}
\right)~,
\end{equation}
where the angular factor, $\Theta_{\varsigma}(\theta,\varphi)$, is
determined by the requirement
$\hat{K}\Theta_{\varsigma}(\theta,\varphi)=\varsigma
\Theta_{\varsigma}(\theta,\varphi)$. This equation for eigenstates
of the angular motion has been investigated by Schr$\ddot{\rm
o}$dinger~\cite{sch38}. He found, the operator $\hat{K}$ is
related to the total angular momentum. Their eigenvalues
satisfy~\cite{gre00}
\begin{equation}
\label{egrelation} \displaystyle
\varsigma=\mp(j+\frac{1}{2})=\left\{
\begin{array}{l}
\displaystyle -(l+1)~~~~{\rm for}~~j=l+\frac{1}{2}
\\[0.2 cm]
\displaystyle l~~~~~~~~~~~~~~{\rm for}~~j=l-\frac{1}{2}
\end{array}
\right.~.
\end{equation}
Also the eigenfunction $\Theta_{\varsigma}$ is related with the
spherical harmonics as follows
\begin{equation}
\label{har} \Theta_{\varsigma}(\theta,\varphi)\propto Y_{l,{\bar
m}+\frac{1}{2}}(\theta,\varphi)\propto P^{{\bar
m}+\frac{1}{2}}_{l}(\cos\theta)e^{i({\bar
m}+\frac{1}{2})\varphi}~,
\end{equation}
or
\begin{equation}
\label{har2} \Theta_{\varsigma}(\theta,\varphi)\propto Y_{l,{\bar
m}-\frac{1}{2}}(\theta,\varphi)\propto P^{{\bar
m}-\frac{1}{2}}_{l}(\cos\theta)e^{i({\bar
m}-\frac{1}{2})\varphi}~,
\end{equation}
where $P(\cos\theta)$ being the associated Legendre functions, and
${\bar m}\pm\frac{1}{2}=0,\pm 1,\pm 2,\cdot\cdot\cdot,\pm l$.

With the help of equations (\ref{Keigen}), (\ref{twocomponent}),
and (\ref{sep}), we can rewrite the equation (\ref{neweq}) into
two equations
\begin{eqnarray}
\label{dir1} && R\left( \frac{\partial}{\partial t }+ i m
 \right)\phi(r,t)=\left[\left(\frac{\partial}{\partial
r}+\frac{1}{r}\right)+\frac{\varsigma}{
r}\right]\tau_{1}\chi(r,t)~,
\\
\label{dir2} && R\left( \frac{\partial}{\partial t }- i m
 \right)\chi(r,t)=\left[\left(\frac{\partial}{\partial
r}+\frac{1}{r}\right)-\frac{\varsigma}{
r}\right]\tau_{1}\phi(r,t)~.
\end{eqnarray}
Here $\frac{1}{r}\tau_{1}\varsigma$ is the term of spin-orbit
coupling. To solve the above equations, we separate the functions
$\phi(t,r)$ and $\chi(t,r)$ into radial and temporal factors,
respectively
\begin{equation}
\label{septr}
\phi(t,r)=U_{1}(r)T_{1}(t)~,~~~~\chi(t,r)=U_{2}(r)T_{2}(t)~.
\end{equation}
With the help of Eq.(\ref{paulima}), inserting the above equations
into Eq.(\ref{dir1}) and Eq.(\ref{dir2}) yields the evolution
equations
\begin{eqnarray}
\label{eveqphi} &&  R\left( \frac{d}{d t }+ i m
 \right)T_{1}(t)-i\kappa T_{2}(t)
=0~,
\\
\label{eveqchi} && R\left( \frac{d}{d t }- i m
 \right)T_{2}(t)-i\kappa T_{1}(t)
=0~,
\end{eqnarray}
and the radial equations
\begin{eqnarray}
\label{spatialeqphi} && \left[\left(\frac{d}{d r}+\frac{1}{
r}\right)+\frac{\varsigma}{ r}\right]\left[\left(\frac{d}{d
r}+\frac{1}{r}\right)-\frac{\varsigma}{
r}\right]U_{1}+\kappa^{2}U_{1}=0~,
\\
\label{spatialeqchi} && \left[\left(\frac{d}{d r}+\frac{1}{
r}\right)-\frac{\varsigma}{ r}\right]\left[\left(\frac{d}{d
r}+\frac{1}{r}\right)+\frac{\varsigma}{
r}\right]U_{2}+\kappa^{2}U_{2}=0~.
\end{eqnarray}
Obviously, if $U_{1}$ is substituted by $U_{2}$, and
simultaneously $\varsigma$ are replaced by $-\varsigma$, then the
equation (\ref{spatialeqphi}) becomes the equation
(\ref{spatialeqchi}). Therefore, in the following, it is of only
necessity to study the equation (\ref{spatialeqphi}).

The equation (\ref{spatialeqphi}) directly simplifies
to~\cite{hua05}
\begin{equation}
\label{simplek0} \frac{d^2 U_{1}}{d r^2}+\frac{2}{r}\frac{d
U_{1}}{d r}+\left(\frac{\varsigma(1-\varsigma)}{ r^2}+\kappa^{2}
\right)U=0~.
\end{equation}
Through a simple transformation, one can easily find that the
equation (\ref{simplek0}) is the Bessel's equation of order
$-\varsigma+\frac{1}{2}$. The solutions of the above equation
are~\cite{wg00}
\begin{equation}
\label{solution00} U_{1}(r)=\frac{1}{\sqrt{\kappa
r}}J_{\pm(\varsigma-\frac{1}{2})}(\kappa r)~.
\end{equation}
Where $J_{\pm(\varsigma-\frac{1}{2})}(\kappa r)$ are called the
Bessel functions. In the case of
$\varsigma=0,-1,-2,\cdot\cdot\cdot$, the Bessel functions in terms
of sine or cosine functions are explicitly taken to be~\cite{wg00}
\begin{equation}\label{Bessel2}
\begin{array}{l}
\displaystyle J_{-\varsigma+\frac{1}{2}}(\kappa
r)=\sqrt{\frac{2}{\pi\kappa r}} \left\{ \sin \left( \kappa
r+\frac{\varsigma \pi}{2} \right)
\sum^{[-\frac{\varsigma}{2}]}_{n=0}
\frac{(-1)^{n}(-\varsigma+2n)!}{(2n)!(-\varsigma-2n)!(2\kappa
r)^{2n}}
\right.\\[1cm]
\displaystyle~~~~~~~~~~~\left.+\cos \left( \kappa
r+\frac{\varsigma \pi}{2} \right)
\sum^{[-\frac{\varsigma+1}{2}]}_{n=0}
\frac{(-1)^{n}(-\varsigma+2n+1)!}{(2n+1)!(-\varsigma-2n-1)!(2\kappa
r)^{2n+1}} \right\} ~,
\end{array}
\end{equation}
and
\begin{equation}\label{Bessel1}
\begin{array}{l}
\displaystyle J_{\varsigma-\frac{1}{2}}(\kappa
r)=\sqrt{\frac{2}{\pi\kappa r}} \left\{ \cos \left( \kappa
r-\frac{\varsigma \pi}{2} \right)
\sum^{[-\frac{\varsigma}{2}]}_{n=0}
\frac{(-1)^{n}(-\varsigma+2n)!}{(2n)!(-\varsigma-2n)!(2\kappa
r)^{2n}}
\right.\\[1cm]
\displaystyle~~~~~~~~~~~\left.-\sin \left( \kappa
r-\frac{\varsigma \pi}{2} \right)
\sum^{[-\frac{\varsigma+1}{2}]}_{n=0}
\frac{(-1)^{n}(-\varsigma+2n+1)!}{(2n+1)!(-\varsigma-2n-1)!(2\kappa
r)^{2n+1}} \right\} ~.
\end{array}
\end{equation}
The formulae for $\varsigma>0$ are easy to read out from above two
equations.

We have indicated that the solution of $U_{2}$ can be obtained
from the solution (\ref{solution00}) by substituting $-\varsigma$
for $\varsigma$. Hence we have acquired the exact forms of the
spatial factor of wave function in de Sitter geometry.

\section{The factor of time evolution}

After setting $s=1/R(t)$, since $R(t)=e^{Ht}$, it is easy to get
the following formula
\begin{equation}
\label{hubble}H=\frac{1}{R}\frac{d R}{d t}=-\frac{1}{s}\frac{d
s}{d t}~.
\end{equation}
In terms of new variable $s$, the evolution equations
(\ref{eveqphi}) and (\ref{eveqchi}) can be rewritten as
\begin{eqnarray}
\label{eveqphis} -H {T}^{\prime}_{1}+\frac{im }{s }T_{1}-i\kappa
T_{2} =0~,
\\
\label{eveqchis} -H {T}^{\prime}_{2}-\frac{im }{s }T_{2}-i\kappa
T_{1} =0~,
\end{eqnarray}
where ${T}^{\prime}\equiv d {{T}}/ds$. From the above equations,
we acquire the equation satisfied by $T_{1}$ as follows
\begin{equation}
\label{eveqt1} s^2 {T}^{\prime\prime}_{1}+\left[ \frac{\kappa^2
s^2 }{H^2 } -\nu(\nu-1)\right] T_{1} =0~,
\end{equation}
where
\begin{equation}
\label{nu} \nu=\frac{im}{H}~~~{\rm or}~~\nu=1-\frac{im}{H}~.
\end{equation}
Obviously, the above equation is the Bessel's equation of order
$\nu-\frac{1}{2}$, the solutions of which is generally presented
as
\begin{equation}
\label{solution1} T_{1}(t)=\frac{1}{\sqrt{R}}
J_{\pm(\nu-\frac{1}{2})}(\frac{\kappa}{HR})~.
\end{equation}
These solutions demonstrate that $\nu=\frac{im}{H}$ and
$\nu=1-\frac{im}{H}$ are quite equivalent. If $2\nu$ isn't an
integer, then the Bessel function $J_{\pm(\nu-\frac{1}{2})}$ can
be explicitly expressed as the infinite series, that is
\begin{eqnarray}
\nonumber J_{\nu-\frac{1}{2}}(\frac{\kappa}{HR}) & = &
\sum^{\infty}_{\iota=0}
\frac{(-1)^{\iota}}{\iota!}\frac{1}{\Gamma(\nu+\iota+\frac{1}{2})}\left(
\frac{\kappa}{2HR}\right)^{2\iota+\nu-\frac{1}{2}}
\\
\label{besselseries} & = &
\left(\frac{\kappa}{2H}\right)^{i\frac{m}{H}}e^{-imt}\sum^{\infty}_{\iota=0}
\frac{(-1)^{\iota}}{\iota!}\frac{1}{\Gamma(\nu+\iota+\frac{1}{2})}\left(
\frac{\kappa}{2HR}\right)^{2\iota-\frac{1}{2}} ~,
\end{eqnarray}
where $\Gamma(\cdot\cdot\cdot)$ is the gamma function. It is easy
to read out the form of $T_{2}(t)$
\begin{equation}
\label{solution2} T_{2}(t)=\frac{1}{\sqrt{R}}
J_{\pm(-i\frac{m}{H}-\frac{1}{2})}(\frac{\kappa}{HR})~.
\end{equation}
Hence we have given the exact solutions of the Dirac equation in
de Sitter space-time.

In a different coordinates of spatially flat Robertson-Walker
space-time, the exact solutions of the Dirac equation with
$R(t)=e^{Ht}$ have been given in Ref.\cite{bd87}. We acquire the
same evolution factor as the result in Ref.\cite{bd87}.

\section{Neutrino oscillations}
Recent neutrino experiments imply that the masses of neutrinos
satisfy $m \sim 10^{-2}$eV~\cite{xin04}. If one select $H$ as the
present-day Hubble constant, then $|\nu|$ is about $10^{30}$.
Therefore, $|\nu|$ is a huge number. Then we can acquire the
approximate formulae of the Bessel function from
Eq.(\ref{besselseries})
\begin{eqnarray}
\nonumber && J_{\nu-\frac{1}{2}}(\frac{\kappa}{HR})\sim
\exp\left\{\nu-\frac{1}{2}+\left(\nu-\frac{1}{2}\right)
\log\left(\frac{\kappa}{2HR}\right)-\nu\log\left(\nu-\frac{1}{2}\right)\right\}
\\
\label{approx}&&~~~~~~~~~~~~~~
\cdot\left[\frac{1}{\sqrt{2\pi}}+\frac{c_{1}}{\nu-\frac{1}{2}}
+\frac{c_{2}}{(\nu-\frac{1}{2})^{2}}+\cdot\cdot\cdot\right]~,
\end{eqnarray}
where $c_{1}$ and $c_{2}$ are small constants. From above
equation, we can reasonably obtain more concise expression as
follows
\begin{eqnarray}
\nonumber &&
 J_{\nu-\frac{1}{2}}(\frac{\kappa}{HR})\sim
\frac{1}{\sqrt{2\pi}} \exp\left\{\nu+\nu\left(
\log\left(\frac{\kappa}{2im}\right)-Ht\right)\right\}
\\
\label{approx2}\displaystyle &&~~~~~~~~~~~~~~ =
\frac{1}{\sqrt{2\pi}}e^{-\frac{3\pi m}{2H}}
\exp\left\{\frac{im}{H}\left[1+\log
\left(\frac{\kappa}{2m}\right)\right]-imt\right\} ~.
\end{eqnarray}
It is well known that the Dirac equation describes both particle
and its antiparticle. So the sine or cosine functions shall be
rewritten as the exponential function with imaginary argument.
According to the exact solutions of the Dirac equation, we can
read out the general form of the phase in de Sitter space-time,
that is
\begin{equation}
\label{phase}{\rm phase}=\frac{m}{H}\left[1+\log
\left(\frac{\kappa}{2m}\right)\right]-mt+\kappa r+({\bar
m}\pm\frac{1}{2})\varphi \pm\frac{\varsigma \pi}{2}
+z_{\varsigma,{\bar m}}(\theta) ~,
\end{equation}
where $z_{\varsigma,{\bar m}}(\theta)$ is a function related with
eigenvalues $\varsigma$, ${\bar m}$ and coordinate $\theta$.

Updated large experiments have shown that neutrinos are possibly
massive although their masses are very tiny. If there are flavor
mixing in neutrino sector, and the leptonic quantum numbers are
only approximately conserved, then there exist the fascinating
possibility of neutrino oscillations.

In the case of three generations of neutrinos, each of the three
neutrino mass eigenstates shall be represented by $\nu_{I}$;
$I=1,2,3$. Because of neutrino mixing, we have the linear
superposition
\begin{equation}
\label{mix}\nu_{\Lambda} =\sum_{I=1,2,3}U_{\Lambda I}\nu_{I}~,
\end{equation}
where $U_{\Lambda I}$ is a unitary matrix, and
$\Lambda=e,\mu,\tau$ represents the weak flavor eigenstates
(corresponding to electron, muon and tau neutrinos respectively).
The $\nu_{1}$, $\nu_{2}$ and $\nu_{3}$ correspond to the three
mass eigenstates of masses $m_1$, $m_2$ and $m_3$, respectively.
Conventionally the mixing matrix $U_{\Lambda I}$ is parameterized
by three mixing angles, one CP violating phase and two Majorana
phases. For the sake of investigation on neutrino oscillations, we
ignore the CP violating phase. In this paper we assume that the
neutrinos are of the Dirac type. Hence, in terms of mixing angles
$\alpha,\beta,\eta$, the mixing matrix reads~\cite{xin04}
\begin{equation}
\label{mixmatrix} U(\alpha,\beta,\eta)=\left(
\begin{array}{ccc}
c_{\alpha}c_{\eta} & s_{\alpha}c_{\eta} & s_{\eta}
\\
-c_{\alpha}s_{\beta}s_{\eta}-s_{\alpha}c_{\beta} &
+c_{\alpha}c_{\beta}-s_{\alpha}s_{\beta}s_{\eta} &
s_{\beta}c_{\eta}
\\
-c_{\alpha}c_{\beta}s_{\eta}+s_{\alpha}s_{\beta} &
-c_{\alpha}s_{\beta}-s_{\alpha}c_{\beta}s_{\eta} &
c_{\beta}c_{\eta}
\end{array}
\right)~,
\end{equation}
where $c_{\xi}\equiv\cos(\xi)$, $s_{\xi}\equiv\sin(\xi)$, with
$\xi=\alpha,\beta,\eta$.

Let us assume: a weak flavor eigenstate, e.g., an electron
neurtino, is emitted at a point $ P$ with coordinates
$P(t_e,r_e,\theta_e,\varphi_e)$. For simplicity, we are interested
in its radial propagation. We also assume that the different mass
eigenstates have the same total angular momentum $\varsigma$. So,
in the momentum representation, the mass eigenstates must have
different eigenvalue $\kappa$, denoted by $\kappa_{I}$. Therefore,
in de Sitter space-time, the radial evolution of them is given by
the expression:
\begin{eqnarray}\displaystyle\nonumber
\nu_{I}(t>t_e,r)=&&\exp\left\{i\frac{m_{I}}{H}\left[1+\log
\left(\frac{\kappa_{I}}{2m_{I}}\right)\right]\right.
\\\displaystyle
\label{ralev} && \left. -im_{I}(t-t_e)+i\kappa_{I}
(r-r_e)+i\Pi\right\}\nu_{I}( P) ~,
\end{eqnarray}
here $\Pi$ being the phase factor which is the same for all mass
eigenstates.

Considering Eq.(\ref{mixmatrix}) and Eq.(\ref{ralev}) together,
one can easily draw the following conclusion: in de Sitter
space-time, under the assumption that the mass eigenstates have
the same total angular momentum, if a weak flavor eigenstate
$\nu_{\Lambda}$ is created at the point $P$, then the probability
to find the weak flavor eigenstate $\nu_{\Lambda^{\prime}}$ at the
point $Q(t_f,r_f,\theta_e,\varphi_e)$ is given by the formulae:
\begin{eqnarray}\displaystyle \nonumber
&& p(\nu_{\Lambda}\to\nu_{\Lambda^{\prime}})=\mid
\nu_{\Lambda^{\prime}}(Q)\mid^{2}
\\
\nonumber  &&=\sum_{I,J=1}^{3}\left\{ U_{\Lambda^{\prime}I}
\exp\left[ -im_{I}(t_f-t_e)+i\kappa_{I}
(r_f-r_e)\right]U_{{\Lambda}I}\right\}
\\
\nonumber &&~~~~\cdot\left\{ U_{\Lambda^{\prime}J}\exp\left[
+im_{J}(t_f-t_e)-i\kappa_{J}
(r_f-r_e)\right]U_{{\Lambda}J}\right\}
\\
\nonumber
&&=\delta_{\Lambda^{\prime}\Lambda}-4U_{\Lambda^{\prime}1}U_{{\Lambda}1}
U_{\Lambda^{\prime}2}U_{{\Lambda}2}\sin^{2}[\omega_{12}]
-4U_{\Lambda^{\prime}1}U_{{\Lambda}1}
U_{\Lambda^{\prime}3}U_{{\Lambda}3}\sin^{2}[\omega_{13}]
\\
\label{probability} &&~~~~-4U_{\Lambda^{\prime}2}U_{{\Lambda}2}
U_{\Lambda^{\prime}3}U_{{\Lambda}3}\sin^{2}[\omega_{23}]~,
\end{eqnarray}
where
\begin{eqnarray}
\label{phasedif1} \omega_{IJ}&=&\frac{\Delta\kappa}{2}\Delta r
-\frac{\Delta m}{2}\Delta t +\frac{\Delta
m}{2H}+\frac{1}{2H}\log\left[
\left(\frac{\kappa_{J}}{2m_{J}}\right)^{m_{J}}
\left(\frac{2m_{I}}{\kappa_{I}}\right)^{m_{I}}\right]~,
\\
\label{phasedif2} \Delta r&=&r_f-r_e~, ~~~~~~~~~~~ \Delta t=
t_f-t_e~,
\\
\label{phasedif3} \Delta\kappa&=&\kappa_{J}-\kappa_{I}~,
~~~~~~~~~~~ \Delta m=m_{J} - m_{I}~.
\end{eqnarray}
We have given the formulae for calculating the neutrino
oscillation probabilities in de Sitter space-time, which
demonstrate that the probabilities are related with not only the
positions, masses and momenta, but also the Hubble constant $H$.
In Minkowski space-time, the Hubble constant doesn't appear.

\section{Conclusion}

The neutrino oscillations in de Sitter space-time is studied
through investigating the covariant Dirac equation. The exact
solutions of the Dirac equation are obtained in the conformally
flat coordinates. The phase of the spin-$\frac{1}{2}$ wave
function is read out. Then the formulae of neutrino oscillation
probabilities in de Sitter space-time are constructed. Our result
is quite different from the stand form in Minkowski space-time
because of the different topology of de Sitter manifold.

{\bf Acknowledgement:}  I am indebted to Prof. Carlo Giunti for
providing me the scanned PDF file of Ref.\cite{sto79} and giving
the helpful comments on an earlier version of this work. I wish to
gratefully thank Prof. D. V. Ahluwalia-Khalilova for his
enlightening communications. I wish to extend my sincere thanks to
Prof. D. Du, Prof. C. J. Zhu, Prof. Z. Chang, Prof. A. Sugamoto
and Prof. I. Oda for their help and encouragement.

\end{document}